\documentclass[conference]{IEEEtran}
\IEEEoverridecommandlockouts
\usepackage{cite}
\usepackage{amsmath,amssymb,amsfonts}
\usepackage{algorithmic}
\usepackage{graphicx}
\usepackage{xcolor}
\usepackage{textcomp}
\usepackage{comment}
\usepackage{placeins}
\usepackage{booktabs}
\usepackage{multirow}
\usepackage{siunitx}
\usepackage{makecell} 
\usepackage{xcolor}

\usepackage{xcolor}
\def\BibTeX{{\rm B\kern-.05em{\sc i\kern-.025em b}\kern-.08em
    T\kern-.1667em\lower.7ex\hbox{E}\kern-.125emX}}
\begin{document}

\title{\textbf{STAR}: Astrocyte-Inspired \underline{ST}ate-\underline{A}ugmented \underline{R}epair for Supervised Memristive AI Hardware Systems
\thanks{Research was sponsored in part by the Army Research Office and was accomplished under Grant Number W911NF-24-1-0127. The views and conclusions contained in this document are those of the authors and should not be interpreted as representing the official policies, either expressed or implied, of the Army Research Office or the U.S. Government. The U.S. Government is authorized to reproduce and distribute reprints for Government purposes notwithstanding any copyright notation herein. This work was also supported in part by the National Science Foundation under award No. CAREER \#2337646. 

The authors acknowledge the use of Claude (Anthropic) in the preparation of this manuscript, limited to language and wording refinement, LaTeX formatting assistance, and generation of the schematic illustration in Fig.~\ref{fig:neuro_experiment}. Claude was not used to generate, analyze, or interpret any technical claims, numerical results, or research data, all of which were produced and independently verified by the authors.
}
}

\author{\IEEEauthorblockN{Yusuf Ahmed Khan, Zhuangyu Han, Abhronil Sengupta}
\IEEEauthorblockA{\textit{School of EECS} \\
\textit{Penn State University}\\
University Park, PA 16802, USA \\
Email: \{yusuf, zfh5141, sengupta\}@psu.edu}

}

\maketitle

\begin{abstract} 

Memristive crossbar arrays have emerged as a promising platform for efficient on-chip learning, enabling local learning rules such as Equilibrium Propagation (EP) to be realized without the memory and data-movement overhead of conventional backpropagation. However, as these devices age, permanent stuck-at (SA) faults accumulate at individual synaptic elements, irreversibly corrupting the stored weights and degrading model performance in ways that in-situ retraining alone cannot address. Existing mitigation strategies either rely on hardware redundancy at significant area and power cost, or require explicit fault localization that is impractical to perform continuously on-chip. This paper adopts a brain-inspired fault recovery route motivated by self-repair functionalities enabled by astrocytes -- a type of glial cell. We couple a computational neuroscience study of astrocytic neuromodulation under permanent synaptic faults with an algorithmic recovery mechanism for faulted crossbar-mapped networks. Our computational neuroscience study characterizes how astrocytes modulate surviving synapses in a bidirectional EP-specific network setting under both SA-0 and high-conductance stuck-at fault conditions, revealing that recovery operates at the neural population level. Motivated by this observation, we propose a retraining-based repair mechanism that augments EP with an additional repair nudge anchored to pre-fault activation targets of the healthy network, encouraging surviving weights to collectively reconstruct pre-fault internal representations without any information of which weights are faulty. To our knowledge, STAR is the first astrocyte-inspired repair mechanism to operate under a supervised local learning rule and to scale to convolutional networks. Experiments across MLP and CNN architectures on MNIST, CIFAR-10, and CIFAR-100 demonstrate recovery gains approaching 60\% over standard in-situ retraining. STAR is further robust to additional hardware non-idealities, including conductance quantization and noisy weight updates, as well as to asymmetric stuck-at fault distributions.

\end{abstract}


\section{Introduction}

Neural network training at scale is dominated by backpropagation (BP) and its temporal variants, including backpropagation through time (BPTT) and recurrent backpropagation (RBP). While effective, these methods carry a fundamental hardware cost: the backward pass requires storing intermediate activations accumulated during the forward computation, imposing significant memory overhead and making fully on-chip learning difficult to realize on resource-constrained platforms. BP is also widely regarded as biologically implausible, as the brain does not appear to propagate global error signals in reverse through its circuits~\cite{b27}.

These limitations have motivated a growing body of work on local learning rules that avoid an explicit backward pass, including Feedback Alignment, Direct Feedback Alignment, Target Propagation, and Predictive Coding~\cite{b28,b29}, demonstrating that competitive accuracy is achievable without weight transport. Motivated by similar biological constraints, Equilibrium Propagation (EP)~\cite{b18,b19} is particularly well-suited to in-memory computing: EP derives weight updates from differences between locally observed neuron states across training phases, requiring neither a stored computational graph nor a dedicated backward pass. Furthermore, EP has been proposed as a biologically motivated alternative to backpropagation, with local co-activation-based updates that are Hebbian in character, placing it in the same family as correlation-based plasticity rules such as Spike-Timing Dependent Plasticity (STDP). This EP locality property maps naturally onto memristive crossbar arrays, where weights are stored and updated directly at the memory element.

However, hardware realization introduces a practical challenge that training algorithms alone cannot address: permanent device faults. As crossbar arrays age, individual synaptic elements develop permanent stuck-at (SA) faults, where a conductance becomes irreversibly fixed at either zero or a saturated high value. Such faults directly corrupt the weight and cannot be corrected by subsequent gradient updates, leading to severe and progressive degradation in model performance. Existing mitigation strategies, discussed further in Section~\ref{sec:related}, either rely on hardware redundancy at significant area and power cost, or require explicit fault localization that is impractical to perform continuously on-chip. The brain, in contrast, contends with synaptic faults yet maintains function through distributed self-repair driven by neuromodulatory signals \cite{b3,b4}.

As neuroscience research progresses, the prevailing neuron-centric view of neural computation is being reconsidered. Glial cells constitute roughly half of all cells in the human brain, with the glia-to-neuron ratio approaching parity in several regions~\cite{b1,b2}, yet their computational roles remain largely absent from artificial neural network models. Among glial cells, astrocytes are of particular interest: beyond providing metabolic support and regulating extracellular ion homeostasis, they actively participate in synaptic modulation, the formation and pruning of synaptic connections, and the broader coordination of neural circuit activity~\cite{b3,b4}. Astrocytes are believed to underlie self-repair functionalities in neural circuits by regulating synaptic transmission probabilities of healthy synapses  through neuromodulatory signaling at tripartite synapses~\cite{b3,b30}, providing a natural biological motivation for augmenting EP training (belonging to the class of STDP-based local learning mechanisms) and corresponding neuromorphic hardware with an astrocyte-inspired autonomous repair mechanism. Our computational neuroscience study (Section~\ref{sec:neuro}), conducted using a bidirectionally coupled neuron-astrocyte network to match EP's recurrent dynamics (EP is applicable to convergent RNNs where every layer is connected to the next by symmetric forward and backward weight connections), reveals that astrocytic modulation acts on the surviving synaptic ensemble rather than targeting individual damaged connections, driving the network back toward its pre-fault population-level activity regime.

\begin{figure}[!t]
    \centering
    \includegraphics[width=0.5\textwidth]{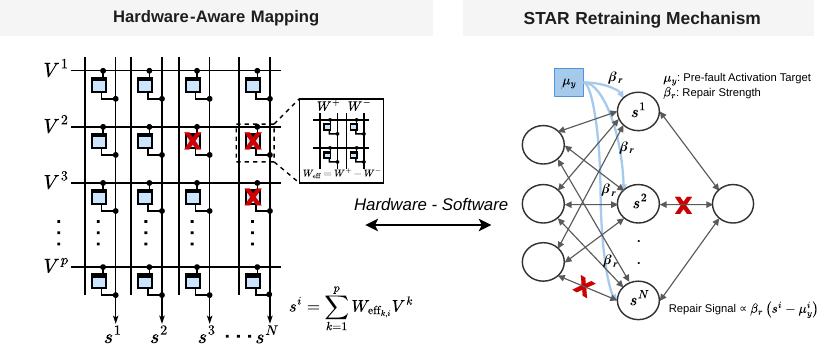}
    \caption{Overview of the STAR framework. (Left) A crossbar array subject to permanent stuck-at faults, where each effective weight $W_{\text{eff}} = W^{+} - W^{-}$ is realized as a differential pair of non-negative memristive conductance branches. The $i$-th output neuron state $s^i$ is computed as $s^i = \sum_{k=1}^{p} W_{\text{eff}_{k,i}} V^k$, where $V^k$ denotes the $k$-th input voltage and $p$ is the number of inputs; the index $i$ is shared with the right panel to make the hardware--software correspondence explicit. (Right) The STAR mechanism applied during the EP nudged phase: prior to fault injection, the pre-fault activation targets $\mu_y$ are recorded from the healthy network; after fault injection, a repair signal proportional to $\beta_r(s^i - \mu_y^i)$ nudges each hidden neuron $s^i$ toward its individual stored activation target $\mu_y^i$ during retraining, without requiring knowledge of which weights are faulty.}
    \label{fig:overview}
\end{figure}

Motivated by this population-level repair principle and by the need for a fault-recovery mechanism compatible with local, on-chip learning, we propose \textbf{STAR: \underline{ST}ate-\underline{A}ugmented \underline{R}epair}. STAR encodes the healthy network's internal representations as compact class-conditional activation targets, then uses these as an additional nudging signal during EP retraining on the faulted network, encouraging the surviving weights to reconstruct the pre-fault equilibrium states without any explicit knowledge of which weights are faulty. Fig.~\ref{fig:overview} provides an overview of this framework, showing both the crossbar hardware mapping and the STAR repair signal applied during retraining.

Our contributions are as follows:
\begin{itemize}
    \item \textbf{Computational Neuroscience Study.} We characterize astrocytic modulation under permanent faults in a bidirectionally coupled neuron-astrocyte network, revealing a population-level recovery principle that motivates our algorithmic design.

    \item \textbf{Physically Realistic Fault Injection.} We construct a hardware-aware fault injection framework mapping EP training onto memristive crossbar arrays, enabling controlled evaluation of both SA-0 and high-conductance stuck-at faults.

    \item \textbf{STAR Fault Recovery Mechanism.} We introduce STAR, which restores network performance after permanent SA fault injection without fault localization. To our knowledge, this is the first astrocyte-inspired repair mechanism formulated under a supervised local learning rule and the first to scale such repair beyond small fully-connected benchmarks (MNIST/FMNIST) to convolutional networks and the CIFAR-10 and CIFAR-100 datasets.

    \item \textbf{Robustness to Hardware Non-Idealities.} We evaluate STAR under 4-bit conductance quantization, noisy weight updates, and asymmetric distributions of SA-0 and high-conductance faults, and show that the repair targets can be stored at a much lower precision than the weights while maintaining recovery.
\end{itemize}

The remainder of this paper is organized as follows. Section~\ref{sec:related} reviews related work. Section~\ref{sec:neuro} presents the neuroscience background and computational neuroscience study. Section~\ref{sec:method} describes the proposed methods. Sections~\ref{sec:exp} and~\ref{sec:result} detail the experimental setup and results. Section~\ref{sec:Discussion} presents the discussions.

\section{Related Works}
\label{sec:related}

\textbf{Hardware Redundancy and Remapping.} Mitigation of permanent SA-0/SA-high faults has traditionally relied on hardware-level redundancy and remapping. Redundancy schemes such as Triple Modular Redundancy (TMR) and Built-In Self-Repair (BISR/BIRA) allocate spare rows, columns, or replicated compute blocks to mask defective units, improving yield at the cost of significant area and power overhead. Mapping and remapping strategies attempt to reduce duplication by identifying faulty cells and permuting weight assignments e.g., crossbar-level remapping, layer ensemble averaging (LEA)~\cite{b10}, or reliability-driven mapping policies in memristive accelerators~\cite{b11,b12}, yet they typically require explicit fault detection and localization and often depend on static fault maps or complex offline solvers.

\textbf{Retraining-Based Recovery.} Retraining-based recovery is attractive as it treats permanent faults as hard constraints and restores task accuracy by re-optimizing only the healthy degrees of freedom, shifting the burden from spare silicon and complex remapping to the model's optimization capacity. In this work, we operationalize this idea using a supervised, local learning rule based on EP, enabling recovery without relying on global backpropagation-style weight transport~\cite{b13}. While conventional fault-aware retraining often depends on global backpropagation and chip-specific fault maps, the proposed approach enables self-repair without additional hardware replication or explicit defect localization, thereby minimizing architectural overhead while restoring application-level accuracy.

\textbf{Astrocyte-Inspired Self-Repair.} Prior astrocyte-inspired self-repair algorithms predominantly operate in unsupervised, STDP-based Spiking Neural Networks on relatively small-scale benchmarks (e.g., MNIST/F-MNIST), often relying on plasticity heuristics (including global statistics such as weight percentiles) to guide recovery after synaptic faults~\cite{b14}, or focusing on local STDP-style updates derived from astrocyte macro-models~\cite{b15}. Separately, coupled spiking astrocyte--neuron systems have demonstrated fault-tolerant maintenance of learned mappings via biologically inspired plasticity rules~\cite{b16,b17}. However, these approaches are typically not formulated under explicit supervised objectives and can face challenges when scaled to deeper supervised architectures due to training complexity. In contrast, STAR targets self-repair in deep supervised networks (including CNNs) under an explicit task objective.

\begin{figure}[!t]
    \centering
    \includegraphics[width=0.7\linewidth]{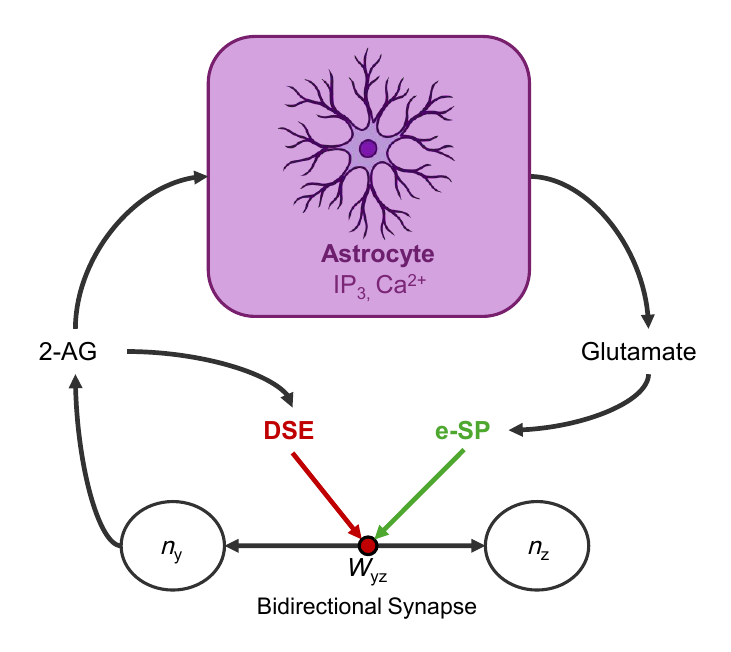}
    \caption{Astrocyte-mediated synaptic modulation under synaptic faults. Neurons $n_y$ (hidden layer) and $n_z$ (output layer) are coupled via bidirectional synaptic weights $W_{yz}$ (mapping to the release probability $\mathrm{PR}(t)$ in Eq.~\eqref{eq:PR}). Neuronal firing in $n_y$ releases 2-AG, which acts on two pathways: directly suppressing synaptic release probability via Depolarization-induced Suppression of Excitation (DSE, red), and triggering intracellular IP$_3$/Ca$^{2+}$ dynamics in the astrocyte, leading to glutamate release and Endocannabinoid-mediated Synaptic Potentiation (eSP, green). Both signals modulate $W_{yz}$, with DSE providing rapid suppression and eSP providing slower distributed potentiation, collectively driving the surviving synaptic population back toward its pre-fault neuronal activity regime.}
    \label{fig:neuro_experiment}
\end{figure}

\section{Computational Neuroscience Modeling}
\label{sec:neuro}

\subsection{Astrocyte-Driven Regulation of Synaptic Efficacy}

Beyond their structural and metabolic roles, astrocytes participate directly in the ongoing regulation of synaptic transmission on timescales longer than individual spikes. Because astrocytes integrate activity across many synapses, they are well positioned to return a slow, modulatory signal that adjusts the probability of neurotransmitter release~\cite{b3,b4,b6}. The net effect on any given synapse reflects a competition between two opposing chemical pathways, one that locally suppresses release probability in the short term, and one that broadly restores or even enhances it over a longer window.

The sequence begins at the postsynaptic membrane, where a firing neuron releases the endocannabinoid 2-arachidonoyl glycerol (2-AG) in a rapid, spike-triggered burst. Between firing events, 2-AG is cleared quickly through enzymatic degradation, so its concentration remains closely tied to recent spiking history \cite{b7}. Formally, this can be expressed as an exponential decay with an impulse production term at each spike time $t_{sp}$:

\begin{equation}
\frac{d(\mathrm{AG})}{dt} = -\frac{\mathrm{AG}}{\tau_{\mathrm{AG}}} + r_{\mathrm{AG}}\,\delta(t - t_{\mathrm{sp}})
\end{equation}

where $\tau_{\mathrm{AG}}$ governs the decay rate and $r_{\mathrm{AG}}$ sets the amount released per spike.

2-AG acts simultaneously on two targets: at the presynaptic terminal, it binds to cannabinoid receptors and directly reduces vesicle release probability, a phenomenon known as depolarization-induced suppression of excitation (DSE). The magnitude of this suppression scales proportionally with the local 2-AG concentration via a coupling constant $K_{\mathrm{AG}}$:

\begin{equation}
\mathrm{DSE} = -\mathrm{AG} \times K_{\mathrm{AG}}
\end{equation}

The second pathway recruits the nearby astrocyte, where 2-AG binding to astrocytic membrane receptors triggers inositol trisphosphate (IP$_3$) production inside the glial cell, driving the release of $\mathrm{Ca}^{2+}$ from the endoplasmic reticulum into the cytoplasm \cite{b8}. The resulting calcium dynamics reflect the interplay of three fluxes: channel-mediated efflux ($J_\mathrm{chan}$), passive leakage from stores ($J_\mathrm{leak}$), and active pump-driven reuptake ($J_\mathrm{pump}$):

\begin{equation}
\frac{d\,\mathrm{Ca}^{2+}}{dt} = J_{\mathrm{chan}} + J_{\mathrm{leak}} - J_{\mathrm{pump}} 
\end{equation}

When cytosolic calcium crosses a threshold level, the astrocyte responds by releasing glutamate through exocytosis. This gliotransmitter spreads across a wider extracellular territory than the original 2-AG signal, reaching multiple nearby synapses \cite{b8}. Its concentration follows the same impulse-decay form as 2-AG, now triggered at calcium threshold-crossing times $t_{\mathrm{Ca}^{2+}}$:

\begin{equation}
\frac{d(\mathrm{Glu})}{dt} = -\frac{\mathrm{Glu}}{\tau_{\mathrm{Glu}}} + r_{\mathrm{Glu}}\,\delta(t - t_{\mathrm{Ca}^{2+}})
\end{equation}

Astrocytic glutamate activates presynaptic metabotropic receptors, inducing a slow, spatially distributed enhancement of release probability termed endocannabinoid-mediated synaptic potentiation (eSP). Unlike DSE, this signal accumulates gradually and persists on a longer timescale $\tau_{\mathrm{eSP}}$, effectively integrating the gliotransmitter input over time:

\begin{equation}
\tau_{\mathrm{eSP}}\frac{d(\mathrm{eSP})}{dt} = -\mathrm{eSP} + m_{\mathrm{eSP}}\,\mathrm{Glu}(t)
\end{equation}

Taken together, these two opposing signals influence the instantaneous release probability $\mathrm{PR}(t)$. Starting from the resting baseline $\mathrm{PR}(0)$, the synapse is continuously pushed and pulled by the sum of the local suppressive term and the distributed potentiating term:

\begin{equation}
\mathrm{PR}(t) = \mathrm{PR}(0) + \mathrm{PR}(0) \times \left(\frac{\mathrm{DSE}(t) + \mathrm{eSP}(t)}{100}\right)
\label{eq:PR}
\end{equation}

This arrangement gives the astrocyte the character of a slow gain controller: brief, intense neuronal activity triggers rapid cannabinoid-mediated suppression first, but sustained or repeated activity eventually recruits enough calcium signalling to tip the balance toward gliotransmitter-mediated potentiation across the broader synaptic ensemble \cite{b9}. While prior works on astromorphic self-repair have already considered computational neuroscience modeling of such astrocyte-neuron-synapse interactions, they primarily consider unidirectional feedforward network settings that do not match the bidirectional weight connectivity associated with EP frameworks. It is therefore necessary to understand and abstract the repair principle for hardware-aware algorithm design  under bidirectional weight coupling.

\subsection{Network Simulations \& Key Insights}
\begin{figure}[!t]
    \centering
    \includegraphics[width=\linewidth]{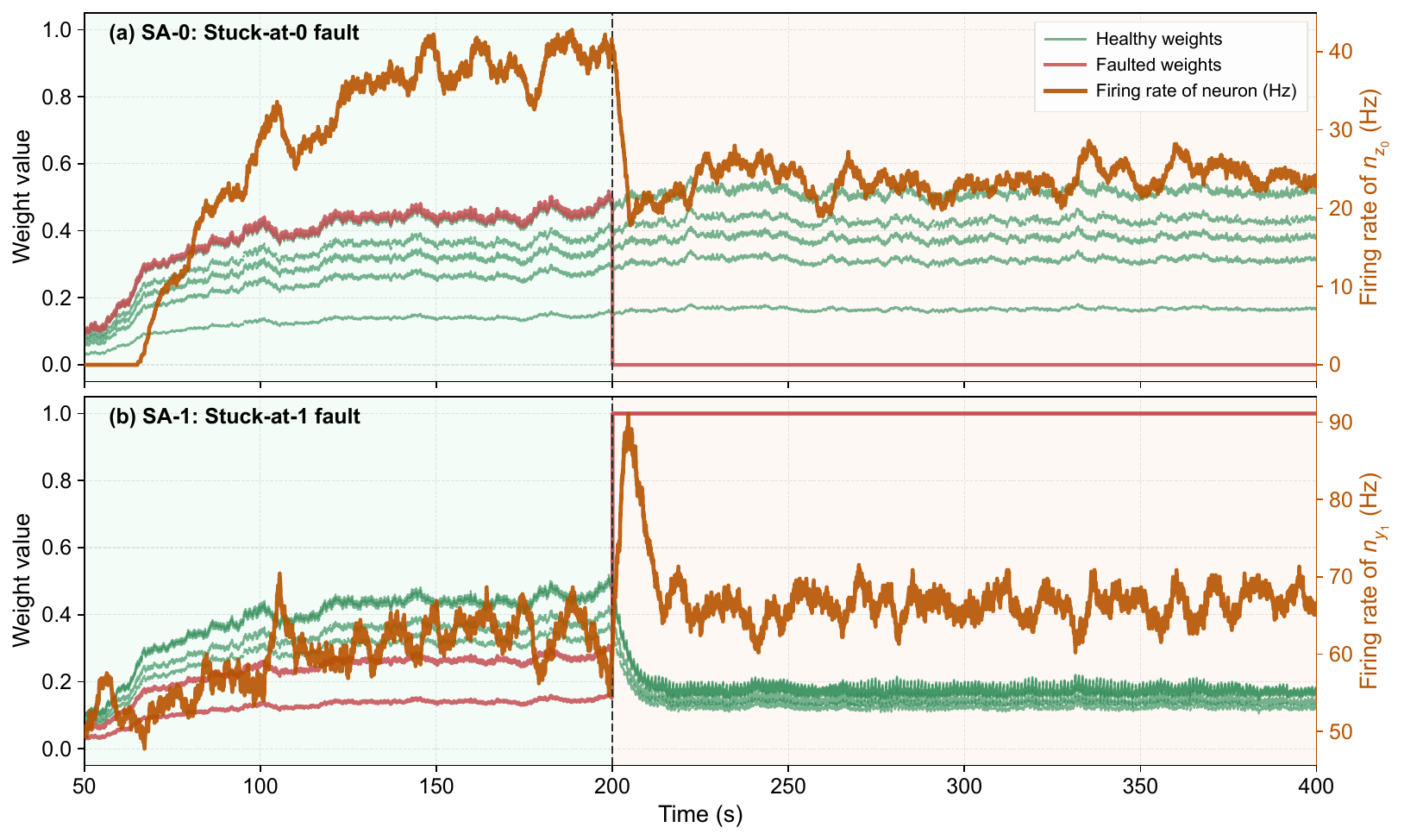}
    \caption{Bidirectional synaptic weight dynamics of the $n_y \rightarrow n_z$ connection under a permanent stuck-at fault injected at $t = 200$\,s. Surviving healthy weights are shown in green, faulted weights in red. (a)~\textbf{SA-0:} the faulted weight is clamped to zero; astrocytic modulation raises the release probability of the surviving synapses, increasing their weights to compensate for the lost conductance. (b)~\textbf{SA-1:} the faulted weight is fixed at maximum conductance; the resulting elevated activity suppresses the surviving synapses, decreasing their weights. In both cases, modulation acts on the surviving population, driving the network back toward its pre-fault activity. The orange curve shows the firing rate of a representative neuron from the affected population (right axis).}
    \label{fig:neuro_graph}
\end{figure}
To characterize the role of astrocytic modulation under permanent synaptic faults, we conducted a computational neuroscience study using a feedforward spiking neural network coupled to a single Li-Rinzel astrocyte model~\cite{b8}. The network comprises an input layer ($n_x = 5$), a hidden layer ($n_y = 3$), and an output layer ($n_z = 2$). The network dynamics are governed by two primary components:

\begin{itemize}
\item \textbf{Neuronal Dynamics:} All neurons are modeled using Leaky Integrate-and-Fire (LIF) dynamics with a membrane time constant ($\tau_{\text{mem}}$) of 9\,ms and a firing threshold of 1.0. The input layer is driven by Poisson-distributed random noise to simulate background neural activity.

\item \textbf{Astrocytic Control:} A single astrocyte monitors the aggregate spiking activity of the hidden layer ($n_y$). This unit calculates intracellular Ca$^{2+}$ dynamics based on neuronal firing rates and generates two competing neuromodulatory signals: DSE and eSP.
\end{itemize}

The weights between layer $n_y$ and $n_z$ are bidirectional ($n_z$ also influences $n_y$ through feedback), consistent with the signalling loop depicted in Fig.~\ref{fig:neuro_experiment}. This design is particularly relevant to EP, which also exerts a bidirectional influence on each layer's neuron state values. The total simulation time was 400\,s. The network scale is intentionally compact: the goal of this study is to characterize the qualitative principle governing astrocytic modulation under bidirectional coupling, namely whether population-level repair acts on surviving synapses, instead of reproducing large-scale spiking dynamics. A minimal network permits unambiguous isolation of the DSE/eSP competition under controlled fault injection, consistent with prior studies in astrocyte-neuron dynamics modeling literature~\cite{b6,b15}.

Stuck-at-0 (SA-0) and high-conductance stuck-at (SA-1) faults were individually simulated within this network. At $t = 200$\,s, each synaptic weight between layers $n_y$ and $n_z$ was permanently set to either zero or one with probability $p_{\text{fault}}$, depending on the fault type under study. Here, the SA-1 condition sets conductance to 1, the maximum value in the normalized weight range of the spiking network, which conceptually parallels the SA-$2\times w_{\max}$ fault in the hardware model of Section~\ref{sec:method}. Under SA-1 faults, the elevated synaptic conductance increased network spiking rate; the astrocyte responded by increasing the magnitude of the DSE signal, reducing the release probability of healthy synapses via Eq.~\eqref{eq:PR} and lowering the average firing rate. Under SA-0 faults, silenced synapses reduced network activity; the astrocyte responded oppositely, reducing the DSE magnitude and raising the release probability of surviving synapses to recover the spiking rate. As shown in Fig.~\ref{fig:neuro_graph}, modulation acts on the surviving synaptic population in both cases, driving the average firing rate of the network toward its pre-fault level. The modulation factor $1 + (\text{DSE}(t) + \text{eSP}(t))/100$, equivalent to $\mathrm{PR}(t)/\mathrm{PR}(0)$ from Eq.~\eqref{eq:PR}, converges as the coupled dynamics reach steady state, causing all surviving synaptic weights to stabilize proportionally at a new equilibrium level regulated by the competition between DSE and eSP.

Mapping this to EP, we can consider the recovery signal to be dependent on the pre-fault neuronal spike rate or activation values. Crucially, EP's nudged phase provides a natural relaxation window: the network is given time to converge to a new steady state after each weight update, and STAR exploits this settling process to steer the hidden layer activations toward the pre-fault targets incrementally across repair epochs. 

\section{Methods}
\label{sec:method}

\subsection{Learning Framework: Equilibrium Propagation}
\label{sec:ep}

EP~\cite{b18,b19,b20} is an energy-based, biologically plausible 
learning rule that computes weight updates purely from local neuron states,
without an explicit backward pass. EP operates on convergent RNNs that receive a static input $x$ and whose
dynamics are governed by a scalar primitive function $\Phi$. The
network state $s$ evolves according to:

\begin{equation}
s^{(t+1)} = \frac{\partial \Phi}{\partial s}(x, s^{(t)}, \theta)
\label{eq:ep_dynamics}
\end{equation}

where $t$ denotes the discrete time step, $\theta$ denotes all
synaptic parameters of the network, including weights and biases, and
$s = \{s^n\}_{n=0}^{L}$ collectively denotes the states of all network
layers. The dynamics are run for $T_{\text{free}}$ steps until convergence to a free-phase
steady state $s^*$, which satisfies $s^* = \partial \Phi/\partial
s|_{s^*}$. For a fully-connected MLP with $L$ layers, the primitive
function is:

\begin{equation}
\Phi(x, s, \theta) = \sum_{n=0}^{L-1} (s^{n+1})^T \cdot w^{n+1} \cdot s^n
\label{eq:phi_mlp}
\end{equation}

where $s^n$ denotes the neuron state vector at layer $n$ and $w^{n+1}$
denotes the weight matrix connecting layers $n$ and $n+1$. For a
convolutional network with $N_{\text{conv}}$ convolutional layers
followed by $N_{\text{fc}}$ fully-connected layers, $\Phi$ extends to:

\begin{equation}
\begin{split}
\Phi(x, s, \theta) =
&\sum_{n \in \text{conv}} s^{n+1} \bullet P(w^{n+1} \star s^n) \\
&+ \sum_{n \in \text{fc}} (s^{n+1})^T \cdot w^{n+1} \cdot s^n
\end{split}
\label{eq:phi_cnn}
\end{equation}

where $\star$ denotes convolution, $P$ a pooling operation, and
$\bullet$ the generalized scalar product over tensors of equal
dimension.

In the nudged phase, run for $T_{\text{nudge}}$ steps, an additional term scaled by $\beta$ is injected into the dynamics, steering the output layer toward the target label $y$:

\begin{equation}
s_\beta^{(t+1)}
=
\frac{\partial \Phi}{\partial s}(x, s_\beta^{(t)}, \theta)
-
\beta \frac{\partial \ell}{\partial s}(s_\beta^{(t)}, y)
\label{eq:nudged_dynamics}
\end{equation}

where $\ell$ is the task loss evaluated at the output layer state (mean squared error for MNIST and CIFAR-10, and cross-entropy with a softmax readout for CIFAR-100), and $\beta$ is the nudging strength. We adopt the 3-phase formulation of~\cite{b19}, which eliminates the first-order bias of the one-sided estimator by running an additional third phase with nudging strength $-\beta$, yielding two nudged steady states $s_{\beta}^*$ and $s_{-\beta}^*$. The weight update follows from the contrastive difference of $\Phi$ evaluated at these two steady states, where $\Delta\theta$ denotes the update to all synaptic parameters $\theta$ (weights and biases, as defined above):

\begin{equation}
\Delta\theta = \frac{\eta}{2\beta}
\left[
\frac{\partial \Phi}{\partial \theta}(x, s_\beta^*, \theta)
-
\frac{\partial \Phi}{\partial \theta}(x, s_{-\beta}^*, \theta)
\right]
\label{eq:weight_update_general}
\end{equation}

where $\eta$ is the learning rate. Since $\Phi$ in Eq.~\eqref{eq:phi_mlp}
is bilinear in the neuron states, the partial derivative with respect
to synapse $\theta_{ij}$ connecting neurons $i$ and $j$ reduces to
the product of their activations at each steady state. Substituting
into Eq.~\eqref{eq:weight_update_general} yields the local contrastive
Hebbian rule:

\begin{equation}
\Delta \theta_{ij} =
\frac{\eta}{2\beta}
\left[
s_{i,\beta}^* s_{j,\beta}^*
-
s_{i,-\beta}^* s_{j,-\beta}^*
\right]
\label{eq:weight_update}
\end{equation}

Each weight update depends only on the activations of the neurons the synapse connects, evaluated at the $+\beta$ and $-\beta$ steady states respectively, with no computational graph required. For convolutional layers, the same local, graph-free contrastive structure is also maintained. EP exhibits conceptual parallels with STDP in that both are local rules modulating synaptic weights based on co-activation of connected neurons; unlike STDP, however, EP operates under an explicit task objective and supports supervised training on deeper architectures, making it well-suited as a scalable learning platform.

\subsection{Hardware-Aware Weight Representation}
\label{sec:hw}

Memristive crossbar arrays implement synaptic dot-product operations by mapping weight values to device conductances. Since physical conductance values are strictly non-negative, signed synaptic weights cannot be represented directly; instead, each effective weight $W_{\text{eff}} = W^+ - W^-$ is computed from two non-negative conductance branches $W^+$ and $W^-$. Biases are implemented analogously, with $b_{\text{eff}} = b^+ - b^-$. In the crossbar implementation, the synaptic parameters $\theta$ introduced in Section~\ref{sec:ep} correspond to these differential conductance pairs $W_{\text{eff}} = W^+ - W^-$, ensuring the EP dynamics operate over a physically realizable weight space.

In addition to permanent faults, practical memristive crossbar deployments are subject to limited conductance precision per cell, variability in the programmed conductance changes, and nonlinear update behavior. The device non-idealities considered in this work are described in Section~\ref{sec:nonideal} and evaluated in Section~\ref{sec:ablation}.

\subsection{Fault Model}
\label{sec:fault}

Permanent stuck-at faults are modeled following prior work~\cite{b15,b10, b21}. Two classes of faults
are injected into the conductance branches $W^+$ and $W^-$ defined in
Section~\ref{sec:hw}:

\begin{itemize}
\item \textbf{Stuck-at-0 (SA-0):} the affected weight is permanently
set to 0, representing a conductance branch that has collapsed to its
off state.
\item \textbf{Stuck-at-$2\times w_{\max}$ (SA-high):} the affected
weight is permanently set to twice the layer-wise maximum conductance
value, representing a high-conductance stuck-at condition. This fault
type has received comparatively less attention in prior astrocyte
self-repair literature, which has predominantly focused on SA-0
conditions~\cite{b15}.
\end{itemize}

The value of $w_{\max}$ is determined through a layer-wise percentile analysis of the trained model weights; further details including the on-off ratio used to derive $w_{\min}$ are provided in Section~\ref{sec:exp}. The two fault types are modeled as equally likely, distributed uniformly across all weights in $W^+$ and $W^-$, with each weight having equal probability $p_{\text{fault}}$ of being affected. We use equal probabilities as the default setting and, since the relative likelihood of SA-0 and SA-high faults depends on device physics, examine asymmetric fault distributions in Section~\ref{sec:ablation}. Faults are permanent, while healthy weights are clamped to the range $[w_{\min}, w_{\max}]$. No fault localization or fault-aware learning signal is used at any point.

\subsection{Device Non-Idealities}
\label{sec:nonideal}

In addition to stuck-at faults, we consider two device non-idealities. First, every conductance branch in $W^+$ and $W^-$ is restricted to 4-bit precision within its layer-wise range $[w_{\min}, w_{\max}]$. While available device programming precision may be lower depending on the particular technology, weight mapping can be easily distributed across multiple devices to effectively implement a higher precision synapse. Stochastic rounding effects are considered in our co-design simulation framework for the quantized programming process to mimic stochasticity involved in device switching. 

We also introduce programming noise by addition of a zero-mean multiplicative Gaussian error with a standard deviation of 25\% of the intended change, drawn independently for every device at every update. This process simulates the effect of device-to-device and cycle-to-cycle variations. Conductance updates are assumed to be linear and symmetric, which prior work has shown to be achievable in memristive synaptic devices~\cite{b31,b32}. In all settings, the faulted devices remain at their stuck values throughout retraining.

\subsection{State-Augmented Repair (STAR)}
\label{sec:star}

STAR is a repair mechanism inspired by the astrocytic principle of maintaining population-level activity, as characterized in our computational neuroscience study (Section~\ref{sec:neuro}). Our analysis follows a four-step deployment pipeline: (1) the model is trained offline on fault-free hardware using EP; (2) the pre-fault activation targets are recorded and stored; (3) permanent stuck-at faults are injected in the deployed crossbar weights; and (4) the faulted model is retrained using STAR nudging to recover performance. Fig.~\ref{fig:overview} illustrates steps (1) and (4) in the context of the crossbar mapping and the repair signal.

This procedure bears a conceptual resemblance to knowledge distillation~\cite{b5}: the pre-fault healthy network plays the role of a teacher, and the faulted network undergoing repair plays the role of a student. Rather than retaining the full teacher model, only compact class-conditional activation statistics are stored and used as fixed guidance signals during retraining, with the formal definition given later. In Section~\ref{sec:ablation}, we further show that these targets can be stored at reduced precision with comparable recovery. A conventional distillation or feature-matching scheme would instead require the healthy teacher network to be retained and evaluated throughout retraining, whereas STAR stores only compact activation targets.

Once SA faults are injected, recovery is achieved through retraining with STAR nudging, which introduces an additional nudging term during the nudged phases of EP training, supplementing the standard task-driven nudge. The STAR term penalizes deviations of the faulted network's neuron states from the pre-fault class-conditional targets, encouraging the surviving weights to restore the internal representations of the healthy model without directly modifying or localizing faulted weights. 
The STAR mechanism integrates seamlessly with the EP framework. EP's nudged phases already provide a principled mechanism to steer neuron states -- STAR adds a secondary nudging term during those phases.

Prior to fault injection, we record the mean free-phase state values by running the trained, fault-free EP model until convergence to the free-phase equilibrium and averaging the equilibrium neuron states across all training samples belonging to each class label. These class-conditional activation targets are recorded once and stored for use during subsequent fault recovery; during retraining, the target corresponding to the ground-truth label $y$ of the current training sample, denoted $\mu_y^n$ for layer $n$, is retrieved. Here $n$ indexes the layer, and the individual component $\mu_y^{n,i}$ denotes the target for neuron $i$ of layer $n$ under the ground-truth label $y$, consistent with the per-neuron repair signal $\beta_r(s^i - \mu_y^i)$ depicted in Fig.~\ref{fig:overview}.

STAR nudging is applied exclusively during the nudged phases (phases 2 and 3 of 3-phase EP). A dedicated repair nudging strength $\beta_r$ controls the influence of the repair signal, independent of the task nudging strength $\beta$. The repair nudging term for the intermediate layers is defined as:
\begin{equation}
R_h(s, y)
=
\frac{1}{N_h}\sum_{n=1}^{N_h} \frac{1}{2}\left\|s^n - \mu_y^n\right\|^2
\label{eq:hidden_penalty}
\end{equation}
where $n$ indexes the intermediate layers, $N_h$ is the total number of intermediate layers, $s^n$ is the neuron state vector of intermediate layer $n$ in the faulted network during retraining, and $\mu_y^n$ is the stored repair target vector for layer $n$ under ground-truth class $y$.

An analogous output-layer repair nudging term $R_{\text{out}}$ is defined for the output neurons using stored target $\mu_y^{N_{out}}$, where $N_{out}$ denotes the output layer index, controlled by a separate coefficient $\beta_{r,\text{out}}$. Unlike the task loss $\ell$, which steers the output toward the ground-truth label, $R_{\text{out}}$ steers the output activations toward the pre-fault activation pattern of the healthy network, encouraging the faulted model to recover its internal output representations independent of the label signal.

These repair nudging terms augment the primitive function $\Phi$, adding two correction terms to the standard nudged primitive:

\begin{equation}
\begin{split}
\Phi_{\text{star}}(x, s, \theta)
=\;& \Phi(x, s, \theta)
-
\beta\,\ell(s, y) \\
&-
\beta_r\,R_h(s, y)
-
\beta_{r,\text{out}}\,R_{\text{out}}(s, y)
\end{split}
\label{eq:phi_aug}
\end{equation}

The augmented neuron update during the nudged phases becomes:

\begin{equation}
s_\beta^{(t+1)}
=
\frac{\partial \Phi_{\text{star}}}{\partial s}(x, s_\beta^{(t)}, \theta)
\label{eq:augmented_state_update}
\end{equation}

The strengths $\beta_r$ and $\beta_{r,\text{out}}$ are hyperparameters tuned relative to $p_{\text{fault}}$, the network architecture, and the dataset. Adding STAR nudging does not alter the structure of the EP weight update: the contrastive update of Eq.~\eqref{eq:weight_update_general} applies with $\Phi_{\text{star}}$ in place of $\Phi$:

\begin{equation}
\Delta \theta = \frac{\eta}{2\beta}
\left[
\frac{\partial \Phi_{\text{star}}}{\partial \theta}(x, s_\beta^*, \theta)
-
\frac{\partial \Phi_{\text{star}}}{\partial \theta}(x, s_{-\beta}^*, \theta)
\right]
\label{eq:star_weight_update}
\end{equation}

This reduces to a local update that depends only on the neuron states in the two augmented steady states $s_\beta^*$ and $s_{-\beta}^*$, preserving the hardware compatibility of the EP weight update. To preserve the first-order bias cancellation of the three-phase symmetric estimator~\cite{b19}, the repair strengths are negated in the third phase alongside the task nudge, i.e., $\beta_r \to -\beta_r$ and $\beta_{r,\text{out}} \to -\beta_{r,\text{out}}$, so that the repair signal is included in $\Phi_{\text{star}}$ symmetrically across both nudged phases. The complete training procedure consists of three phases: Phase~1 (free phase) establishes $s^*$; Phases~2 and~3 run the augmented dynamics with $+\beta$ and $-\beta$ respectively, each incorporating the STAR repair nudge; and the weight update follows from Eq.~\eqref{eq:star_weight_update}.

\subsection{Convergence Property of EP with STAR}
\label{sec:convergence}

STAR augments the primitive function with the repair nudging terms $R_h$ and $R_{\text{out}}$, both of which are quadratic and hence continuously differentiable. Since the base EP framework already guarantees fixed-point convergence under Lipschitz-continuous dynamics~\cite{b18,b20}, and since the sum of Lipschitz-continuous functions is itself Lipschitz continuous, the augmented transition function $F_{\text{star}}(s^{(t)}) = \partial \Phi_{\text{star}} / \partial s$ remains Lipschitz continuous~\cite{b18}. Extending the argument of~\cite{b5} to include the two independent repair terms $R_h$ and $R_{\text{out}}$, the neuron dynamics under $\Phi_{\text{star}}$ converge to a fixed point, provided the hyperparameters satisfy:

\begin{equation}
K_1 + |\beta| K_2 + |\beta_r| K_3 + |\beta_{r,\text{out}}| K_4 < 1
\end{equation}

where $K_1$ bounds the Lipschitz constant of the base transition function, and $K_2$, $K_3$, $K_4$ bound those of the task loss, intermediate repair nudging term, and output repair nudging term respectively. Because the added repair terms $R_h$ and $R_{\text{out}}$ are quadratic, their Lipschitz constants $K_3$ and $K_4$ are finite, so the contraction condition above can be satisfied and a fixed point is guaranteed to exist. The EP gradient theorem continues to hold at the STAR-perturbed steady states. The sensitivity of repair quality to $\beta_r$, together with convergence over an extended retraining period, is examined empirically in Section~\ref{sec:result}.

\begin{table*}[!t]
\centering
\caption{Comparison of baseline retraining (no STAR) and STAR performance under varying
fault probabilities ($p_{\text{fault}}$) }
\label{tab:baseline_vs_star}
\small
\setlength{\tabcolsep}{6pt}
\renewcommand{\arraystretch}{1.2}
\begin{tabular}{l l c c c c c}
\toprule
\hline
\textbf{Model} & \textbf{Dataset} & \textbf{Clean ($\%$)} & $p_{\text{fault}}$ & \textbf{No STAR ($\%$)} & \textbf{STAR ($\%$)} & $\Delta ($\%$)$ \\
\midrule
\multirow{5}{*}{MLP-1H}
& \multirow{5}{*}{MNIST}
& \multirow{5}{*}{97.95}
& 0.5 & $68.49 \pm 5.85$  & $\mathbf{94.10 \pm 0.44}$ & $+25.61$ \\
& & & 0.6 & $48.70 \pm 15.53$ & $\mathbf{93.55 \pm 0.34}$ & $+44.86$ \\
& & & 0.7 & $38.99 \pm 11.68$ & $\mathbf{91.39 \pm 0.48}$ & $+52.41$ \\
& & & 0.8 & $34.99 \pm 7.19$  & $\mathbf{85.43 \pm 0.75}$ & $+50.44$ \\
& & & 0.9 & $25.68 \pm 4.72$  & $\mathbf{73.21 \pm 1.39}$ & $+47.53$ \\
\midrule
\multirow{5}{*}{MLP-2H}
& \multirow{5}{*}{MNIST}
& \multirow{5}{*}{97.42}
& 0.5 & $42.65 \pm 18.60$ & $\mathbf{82.99 \pm 1.37}$ & $+40.34$ \\
& & & 0.6 & $41.94 \pm 19.94$ & $\mathbf{78.83 \pm 1.40}$ & $+36.88$ \\
& & & 0.7 & $33.77 \pm 13.66$ & $\mathbf{71.68 \pm 2.70}$ & $+37.91$ \\
& & & 0.8 & $19.06 \pm 12.12$ & $\mathbf{62.64 \pm 2.65}$ & $+43.58$ \\
& & & 0.9 & $15.55 \pm 7.55$  & $\mathbf{49.31 \pm 2.37}$ & $+33.76$ \\
\midrule
\multirow{5}{*}{VGG-5}
& \multirow{5}{*}{CIFAR-10}
& \multirow{5}{*}{87.42}
& 0.09 & $33.07 \pm 28.13$ & $\mathbf{75.86 \pm 1.30}$ & $+42.78$ \\
& & & 0.10 & $28.42 \pm 19.83$ & $\mathbf{72.31 \pm 2.48}$ & $+43.88$ \\
& & & 0.11 & $31.77 \pm 22.30$ & $\mathbf{73.77 \pm 2.00}$ & $+41.99$ \\
& & & 0.12 & $13.34 \pm 6.66$  & $\mathbf{72.65 \pm 1.62}$ & $+59.31$ \\
& & & 0.13 & $21.89 \pm 23.78$ & $\mathbf{69.97 \pm 2.09}$ & $+48.08$ \\
\midrule
\multirow{5}{*}{VGG-5}
& \multirow{5}{*}{CIFAR-100}
& \multirow{5}{*}{66.38}
& 0.08 & $44.05 \pm 24.32$ & $\mathbf{63.30 \pm 0.11}$ & $+19.25$ \\
& & & 0.09 & $29.37 \pm 25.69$ & $\mathbf{62.54 \pm 0.42}$ & $+33.17$ \\
& & & 0.10 & $12.88 \pm 16.48$ & $\mathbf{59.73 \pm 2.06}$ & $+46.85$ \\
& & & 0.11 & $14.94 \pm 20.65$ & $\mathbf{56.86 \pm 3.12}$ & $+41.92$ \\
& & & 0.12 & $8.29 \pm 13.21$  & $\mathbf{52.43 \pm 2.76}$ & $+44.14$ \\
\hline
\bottomrule
\end{tabular}
\par\vspace{1mm}
\begin{minipage}{0.9\linewidth}
\footnotesize
\emph{Clean}: fault-free software accuracy. \emph{No STAR}: accuracy with EP retraining under faults with the STAR repair signal disabled ($\beta_r = \beta_{r,\text{out}} = 0$).
\end{minipage}
\end{table*}


\begin{table}[!t]
\caption{Training hyperparameters}
\label{tab:hyperparams}
\centering
\footnotesize
\setlength{\tabcolsep}{2.5pt}
\begin{tabular}{lcccc}
\toprule
\hline
\textbf{Param} & \textbf{\makecell{MNIST\\(1H)}} & \textbf{\makecell{MNIST\\(2H)}} & \textbf{CIFAR-10} & \textbf{CIFAR-100} \\
\midrule
\multicolumn{5}{c}{\textbf{Optimization}} \\
\midrule
Activation & Steep.$^*$ & Steep.$^*$ & Hard$^\dagger$ & Hard$^\dagger$ \\
$\beta$    & 0.1 & 0.5 & 0.5 & 0.5 \\
Learning Rate & \makecell{0.25, 0.15,\\ 0.1} & \makecell{0.2, 0.1,\\ 0.05} & \makecell{0.25, 0.15,\\ 0.1, 0.08,\\ 0.05} & \makecell{0.0625, 0.0375,\\ 0.025, 0.02,\\ 0.0125} \\
\midrule
\multicolumn{5}{c}{\textbf{EP Dynamics}} \\
\midrule
$T_{\text{free}}$  & 30 & 100 & 250 & 250 \\
$T_{\text{nudge}}$ & 10 & 20  & 30  & 30 \\
Batch Size         & 20 & 20  & 128 & 128 \\
Epochs             & 30 & 50  & 120 & 150 \\
\midrule
\multicolumn{5}{c}{\textbf{STAR Retraining}} \\
\midrule
$\beta_r,\,\beta_{r,\text{out}}$ & 4,\;4 & 4,\;4 & 1,\;0 & 5,\;0 \\
Retraining Epochs & 1 & 1 & 7 & 20 \\
\hline
\bottomrule
\multicolumn{5}{l}{%
  $^*$Steepened sigmoid: $\sigma(x)=(1+e^{-4(x-0.5)})^{-1}$.} \\
\multicolumn{5}{l}{%
  $^\dagger$Hard sigmoid: $\sigma(x)=(1+\mathrm{htanh}(x-1))/2$.} \\
\end{tabular}
\end{table}

\section{Experimental Setup}
\label{sec:exp}

\subsection{Model Architecture and Datasets}

We evaluate two architecture families: MLP-based and CNN-based. The MLPs use 1 and 2 hidden layers; the CNN is a VGG-5-like network with 4 convolutional layers and 1 fully connected layer. For evaluating the effectiveness of our proposed method, we conducted evaluation on the MNIST, CIFAR-10, and CIFAR-100 benchmark datasets. 

\subsection{Faults and Baselines}
As described in Section~\ref{sec:method}, two fault types are injected: SA-0 and SA-$2{\times}w_{\max}$, with healthy weights clamped to $[w_{\min}, w_{\max}]$. The value of $w_{\max}$ is determined through a layer-wise percentile analysis of the stored model weights. The corresponding $w_{\min}$ is derived by fixing the on-off ratio $w_{\max}/w_{\min}$ to 100.

For MLP architectures, fault rates ranging from 50\% to 90\% were evaluated. For the CNN architecture, faults were injected at rates up to 13\%, reflecting the considerably greater sensitivity of convolutional networks to permanent weight faults relative to fully connected architectures. This heightened sensitivity arises from the CNN's shared-weight kernel architecture, where a single faulty weight, whether stuck-at-0 or stuck-at-$2{\times}w_{\max}$, is reused across every spatial position of the input rather than affecting a single connection as in an MLP, systematically corrupting the entire output feature map. The situation also translates to memristive hardware accelerators by the static nature of kernel weights: in memristor-array-based CNN accelerators operating under weight-stationary dataflow, fixed memristive crosspoint devices store kernel weights that are reused across all spatial positions during inference, meaning a permanent stuck-at fault in any such device poisons every computation that utilizes it~\cite{b21,b22}. Prior work has shown that even a single bit-flip in a CNN weight can cause a dramatic drop in classification accuracy~\cite{b23}, a sensitivity not typically observed at comparable fault rates in fully connected architectures~\cite{b24}.

Faults were injected into pre-trained models, with training hyperparameters detailed in Table~\ref{tab:hyperparams}. 
The baseline for comparison is retraining under faults without any STAR nudging, i.e., with $\beta_r = \beta_{r,\text{out}} = 0$. Given the difference in model complexity and fault severity, retraining was performed for 1 epoch for MLP architectures, and for 7 and 20 epochs for the CNN on CIFAR-10 and CIFAR-100, respectively. Since the CIFAR-100 output is a softmax readout rather than an output neuron layer, the output repair term is not applied ($\beta_{r,\text{out}} = 0$). The underlying hyperparameters for EP optimization and dynamics follow the 3-phase EP formulation of~\cite{b19}, with architecture-specific adjustments; repair hyperparameters $\beta_r$ and $\beta_{r,\text{out}}$ were tuned independently per architecture. All reported results have been averaged over 5 seed runs.

Prior astrocyte-inspired self-repair methods operate exclusively in unsupervised, STDP-driven spiking networks without an explicit task loss~\cite{b14,b15,b16,b17}; the training paradigms, plasticity rules, and evaluation protocols differ fundamentally from the supervised EP setting studied here, rendering direct numerical comparison methodologically inappropriate. The natural baseline within the same learning framework is therefore unassisted EP retraining, which isolates the contribution of the STAR repair signal from any differences attributable to learning rule or task formulation.

\section{Results}
\label{sec:result}

\subsection{Recovery under Faults}

\begin{figure*}[!t]
    \centering
    \includegraphics[width=\textwidth]{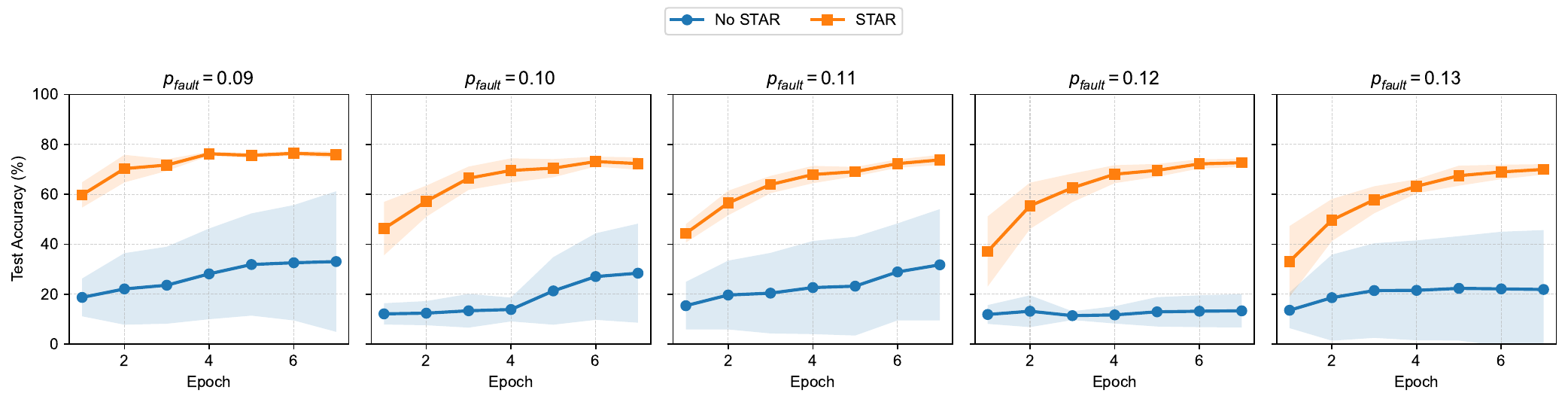}
    \caption{Epoch-wise test accuracy across retraining epochs for the VGG-5 CNN on CIFAR-10, expanding upon the converged results reported in Table~\ref{tab:baseline_vs_star} across fault probabilities ($p_{\text{fault}}\!=\!0.09$--$0.13$); shaded regions denote standard deviation. While No STAR retraining collapses or remains severely degraded at higher fault rates, STAR exhibits consistent recovery across epochs, demonstrating robust fault tolerance even in regimes where standard retraining fails.}
    \label{fig:CNN_results}
\end{figure*}

As the fault rate increases, EP without STAR exhibits rapid performance degradation, as reported in Table~\ref{tab:baseline_vs_star} and illustrated in Fig.~\ref{fig:CNN_results}. With STAR active, the additional repair nudging consistently mitigates the impact of stuck-at faults across all tested fault rates. The repair hyperparameters $\beta_r$ and $\beta_{r,\text{out}}$ were tuned independently for each architecture and dataset to maximize recovery.

For MLP-1H, STAR achieves gains of up to $+52.41$\% over unassisted retraining. Recovery remains consistent for MLP-2H, confirming the mechanism scales to fully connected architectures. For the CNN on CIFAR-10, STAR stabilizes accuracy in the range of $69.97$--$75.86$\% against a baseline that frequently collapses entirely. On CIFAR-100, STAR maintains accuracy in the range of $52.43$--$63.30$\% across $p_{\text{fault}} = 0.08$--$0.12$, and outperforms unassisted retraining for every fault seed at every fault rate. In a fault-free control under the same retraining schedule, STAR and unassisted retraining reach $66.18$\% and $66.50$\% respectively, indicating that the repair signal does not impair a healthy network.

The wide spread in accuracies without STAR reflects the seed-dependent nature of recovery: depending on the fault sampling seed, the network either retains usable accuracy or collapses almost completely. In contrast, STAR consistently anchors recovery toward the pre-fault representations, resulting in substantially lower variance across seeds and confirming that the repair mechanism stabilizes the retraining process under fault stochasticity.

\subsection{Ablation and Robustness Studies}
\label{sec:ablation}

\begin{figure}[!t]
    \centering
    \includegraphics[width=0.8\linewidth]{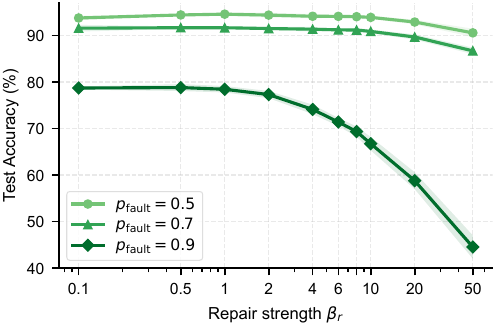}
    \caption{Test accuracy of MLP-1H as a function of repair strength
$\beta_r$ ($\beta_{r,\mathrm{out}} = 4$) across three fault rates;
shaded regions denote standard deviation.
At $p_{\text{fault}} \in \{0.5, 0.7\}$, recovery remains stable
across $\beta_r \in [0.5, 10]$, indicating low sensitivity to precise
tuning. At $p_{\text{fault}} = 0.9$, the optimal $\beta_r$ shifts
toward smaller values, suggesting that fault-rate-aware tuning could
yield further gains at extreme fault rates. The results reported in
Table~\ref{tab:baseline_vs_star} use $\beta_r = 4$ uniformly across
all fault rates.}
    \label{fig:ablation}
\end{figure}

\textbf{Sensitivity to repair strength.} As shown in Fig.~\ref{fig:ablation}, recovery quality is largely insensitive to $\beta_r$ across a wide range at low-to-moderate fault rates. For $p_{\text{fault}} \in \{0.5, 0.7\}$, test accuracy remains stable across $\beta_r \in [0.5, 10]$, varying by less than 1\%, indicating that the mechanism does not require careful hyperparameter tuning to operate effectively in these regimes. At $p_{\text{fault}} = 0.9$, however, the optimal $\beta_r$ shifts toward smaller values, with peak recovery occurring around $\beta_r \in [0.5, 1]$ and degrading monotonically at larger values. Thus at extreme fault rates, larger repair signals tend to degrade rather than improve recovery. The results, reported in Table~\ref{tab:baseline_vs_star}, use $\beta_r = 4$ uniformly across all fault rates; fault-rate-aware tuning of $\beta_r$ represents a promising direction for further improvement, particularly at high fault rates.

\begin{figure}[!t]
    \centering
        \includegraphics[width=0.8\linewidth]{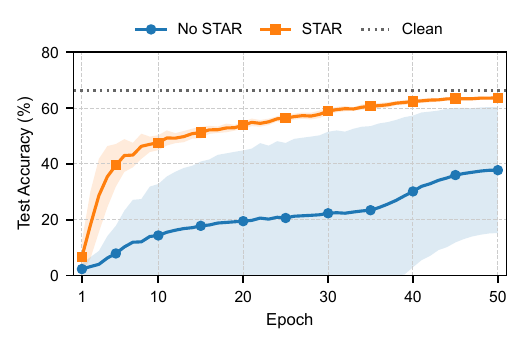}
    \caption{Test accuracy of the VGG-5 CNN on CIFAR-100 over 50 retraining epochs at $p_{\text{fault}} = 0.10$; shaded regions denote standard deviation, and the dotted line marks the fault-free accuracy. STAR converges to within $2.7$\% of the clean accuracy, while unassisted retraining remains strongly dependent on the fault seed.}
    \label{fig:convergence}
\end{figure}

\textbf{Extended convergence.} To assess convergence over a longer retraining period, we extended CIFAR-100 retraining at $p_{\text{fault}} = 0.10$ to 50 epochs, with the learning rate schedule extended accordingly (Fig.~\ref{fig:convergence}). STAR converges to $63.69 \pm 0.47$\%, within $2.7$\% of the clean accuracy, whereas unassisted retraining ends at $37.76 \pm 22.53$\% with high variance across fault seeds, several of which remain near chance accuracy for most of the retraining period.

\begin{figure}[!t]
    \centering
        \includegraphics[width=0.8\linewidth]{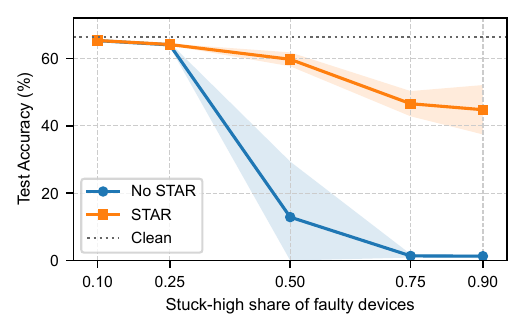}
    \caption{Test accuracy on CIFAR-100 as a function of the fraction of SA-high faults ($p_{\text{fault}} = 0.10$); shaded regions denote standard deviation. Unassisted retraining fails once SA-high faults dominate, whereas STAR maintains recovery across all fault distributions.}
    \label{fig:sa_ratio}
\end{figure}

\textbf{Asymmetric stuck-at fault distributions.} To study the effect of the relative likelihood of the two fault types, we varied the fraction of SA-high faults from 0.10 to 0.90 at $p_{\text{fault}} = 0.10$ on CIFAR-100 (Fig.~\ref{fig:sa_ratio}). Increasing the fraction of SA-high faults makes recovery progressively harder, and once SA-high faults account for half or more of all faults, unassisted retraining largely fails while STAR continues to recover, retaining $44.76$\% accuracy even at a fraction of 0.90.

\begin{table}[!t]
\centering
\caption{Recovery under device non-idealities on CIFAR-100 ($p_{\text{fault}} = 0.10$)}
\label{tab:nonideal}
\resizebox{\columnwidth}{!}{%
\begin{tabular}{lcccc}
\toprule
\hline
\textbf{Setting} & \textbf{Clean ($\%$)} & \textbf{No STAR ($\%$)} & \textbf{STAR ($\%$)} & $\Delta ($\%$)$ \\
\midrule
Full precision & 66.38 & $12.88 \pm 16.48$ & $\mathbf{59.73 \pm 2.06}$ & $+46.85$ \\
Noisy updates (25\%) & 66.38 & $18.23 \pm 20.10$ & $\mathbf{61.58 \pm 0.49}$ & $+43.35$ \\
4-bit weights & 64.48 & $7.04 \pm 13.51$ & $\mathbf{50.70 \pm 1.84}$ & $+43.66$ \\
4-bit + Noisy updates (25\%) & 64.48 & $5.87 \pm 10.89$ & $\mathbf{50.51 \pm 2.40}$ & $+44.64$ \\
\hline
\bottomrule
\end{tabular}}
\par\vspace{1mm}
\footnotesize \emph{Clean}: fault-free accuracy of the starting model (for 4-bit weights, after quantization-aware fine-tuning).
\end{table}

\textbf{Noisy weight updates.} With the noisy updates described in Section~\ref{sec:nonideal} at $p_{\text{fault}} = 0.10$ on CIFAR-100, the recovery provided by STAR is not degraded by update noise, while unassisted retraining remains strongly dependent on the fault seed (Table~\ref{tab:nonideal}). In the corresponding fault-free controls, unassisted retraining and STAR reach $66.27$\% and $65.83$\% respectively.

\begin{table}[!t]
\centering
\caption{STAR recovery and storage for reduced-precision targets on CIFAR-100 (4-bit weights, $p_{\text{fault}} = 0.10$)}
\label{tab:target_precision}
\resizebox{\columnwidth}{!}{%
\begin{tabular}{lccc}
\toprule
\hline
\textbf{Target precision} & \textbf{Target storage (MB)} & \textbf{Weights / targets} & \textbf{STAR ($\%$)} \\
\midrule
2-bit & 1.45 & $2.69\times$ & $51.24 \pm 3.06$ \\
1-bit & 0.72 & $5.38\times$ & $47.21 \pm 4.93$ \\
\hline
\bottomrule
\end{tabular}}
\par\vspace{1mm}
\footnotesize With full-precision targets ($23.14$\,MB), STAR reaches $50.70 \pm 1.84\%$ accuracy.
\end{table}

\textbf{Conductance quantization and target storage.} For the 4-bit evaluation, we first obtain a 4-bit model through quantization-aware fine-tuning of the full-precision network, which reaches $64.48$\%, after which faults are injected at $p_{\text{fault}} = 0.10$ and retraining proceeds with 4-bit weights. STAR retains a large margin over unassisted retraining, which falls below 5\% accuracy in four of the five fault seeds (Table~\ref{tab:nonideal}). The standalone ablation with just quantized states demonstrates that the repaired accuracy degradation is primarily due to the quantization effect and can be potentially eliminated by mapping multiple devices to a singular synaptic state. 

Since the weights are mapped to device conductances while the STAR targets would be held in on-chip buffers, the precision of the stored targets determines the storage cost of repair. 
For CIFAR-100, storing the targets at 2 and 1 bits makes them $2.7\times$ and $5.4\times$ smaller than the $3.9$\,MB occupied by the 4-bit weights (Table~\ref{tab:target_precision}). 2-bit targets give recovery comparable to full-precision targets, and 1-bit targets, which only retain whether each neuron is active for a given class, lead to a moderate reduction, while both remain well above unassisted retraining. Since conductance-mapped weights and buffer-resident targets differ in their physical implementation, this comparison is intended as a qualitative estimate of the relative storage cost.


\FloatBarrier
\section{Discussions}
\label{sec:Discussion}
This work addresses a fundamental reliability challenge in AI hardware accelerators: permanent stuck-at faults in memristive crossbar arrays, which corrupt weights irreversibly and cannot be resolved through continued retraining alone. 
To our knowledge, STAR is the first astrocyte-inspired repair mechanism formulated under a supervised local learning rule, and the first to scale such repair to convolutional architectures and the CIFAR-10 and CIFAR-100 datasets, where it delivers consistent recovery with substantially reduced variance even at fault rates that collapse unassisted retraining. STAR is further robust to additional hardware non-idealities, including 4-bit conductance quantization and noisy weight updates, under which it retains a large margin over unassisted retraining. Its benefit also grows as high-conductance faults dominate the fault distribution. Because it operates purely at the algorithmic level, STAR is complementary to circuit-level redundancy schemes rather than a replacement for them. Several directions remain open: fault-rate-adaptive tuning of $\beta_r$ to improve recovery at extreme fault rates, a quantitative mapping between the astrocytic dynamics and the STAR repair term, evaluation under nonlinear conductance updates, conductance drift, and sub-4-bit precision, scaling STAR to deeper networks and larger-scale tasks, validation on physical crossbar hardware, and a deeper exploration of astrocytic computational contributions, drawing further on principles from neuro-inspired AI.

\end{document}